\documentclass[conference]{IEEEtran}
\IEEEoverridecommandlockouts
\usepackage{url}
\usepackage{cite}
\usepackage{amsmath,amssymb,amsfonts}
\usepackage{algorithmic}
\usepackage{graphicx}
\usepackage{textcomp}
\usepackage{xcolor}
\usepackage{graphicx} 
\usepackage{epstopdf}
\usepackage{float}
\usepackage{subfigure}
\usepackage{array}
\usepackage{booktabs} 
\usepackage{multirow}
\usepackage[]{caption2} 
\def\BibTeX{{\rm B\kern-.05em{\sc i\kern-.025em b}\kern-.08em
    T\kern-.1667em\lower.7ex\hbox{E}\kern-.125emX}}
\begin{document}

\title{Traffic-aware Threshold Adjustment for NFV Scaling using DRL\\}
\author{\IEEEauthorblockN{Hua Chai}
	\IEEEauthorblockA{ State Key Laboratory of Networking and Switching Technology, BUPT}
\{huachai, jiaozhang\}@bupt.edu.cn\\
}
\maketitle

\begin{abstract}
This document is a model and instructions for \LaTeX.
\end{abstract}


\section{Introduction}

Traditional hardware-based network functions are notoriously hard and costly to deploy and scale. The emerging paradigm of network function virtualization (NFV) advocates deploying software network functions in virtualized environments (e.g., VMs) on run on top of general purpose hardware, to significantly simplify deployment and scaling at much lowered costs, and delivering different sets of service to different traffic. 

According to recent authoritative reports \cite{delimitrou2014quasar}, the resource utilization ratio of modern data centers is only 15\%-30\%, which hurts both cost effectiveness and future scalability. Such current situation is caused by provider do not timely adjust resource orchestration according to workload dynamics and physical resource requirements of data centers. A good way to improve current situation is allocating resources on demand. NFV scaling is just a good example, provider can dynamically grow or shrink its capacity by spinning up or spinning down VNF instances. Specifically, a user could request higher capacity ahead of time in order to deal with a known increase in demand in the future, and the provider also can reduce capacity as the traffic volume decreases. The operations and management system could initiate scaling to maintain service level requirements based on monitoring status metrics; or the deployed service components themselves could manage their resource needs similarly to automatic multi-threading scaling of some software running on multi-core CPUs\cite{szabo2015elastic}.
    
The possibility to dynamically scale network services at run-time in an automated fashion is one of the main advantages offered by the NFV approach, providing both better resource utilization and better service at a lower cost\cite{szabo2015elastic}. NFV enables dynamic scale in for energy efficiency \cite{mathew2012energy}, scale out for Service Level Agreement (SLA) management \cite{Cloudward b},\cite{fan2015grep}, and load balancing. 

However, satisfying the QoS of network service while still keeping cost low is challenging for Telco systems due to the variety of service requests, the workload mix, and internal application phases and changes. Therefore, resource allocation is very challenging. Over-provisions leads the cloud service provider to pay for the wasted resources. Under-provisions are much worse, a VNF instance cannot be available immediately \cite{yi2016dynamic}, since VM setup and VNF initiate also need time.The lag time could be as long as 10 min or more to start an instance in Microsoft Azure and the lag time could be various from time to time\cite{hill2010early}, therefore, under-provision may interrupt service, and cause significant SLA violations.   

Both industry and academia have drawn significant attention in NFV scaling. Despite their wide diversity and hybridizations of two or more methods, we classify auto-scaling techniques into reactive and proactive schemes considering the anticipation capacity as the main criteria. As the typical example of reactive scaling, threshold-based rules or policies are very popular among cloud providers like AWS \cite{bworld} and Windows Azure \cite{bworld2}, on account of the simplicity and intuitive nature of these policies. However, selection of thresholds is tricky and it is unable to cope with sudden workload bursts. A desirable solution would require an ability to promptly allocate resource close to the workload. 
  The emergence of dynamic thresholds has improved some of the shortcomings of threshold-based schemes. For example, in \cite{lim2009automated} \cite{beloglazov2010adaptive}, they all adjust the threshold based on the resource usage of the VNF observed in each cycle. \cite{lim2009automated} is using a proportional integrator in cybernetics to set thresholds. This dynamic scaling improves the limitations of static thresholds to a certain extent, but their granularity is still relatively coarse, because all VMs have the same threshold in spite of their original diversities, which does not apply to the actual scenario.
Anyway, reactive policy has its inherent limitations, they cannot allocate resources ahead of time, so the effect of a scaling-up action might arrive too late because adding a new VM in real cloud providers might take up to 15 minutes.

Relatively, the proactive solutions have better ability to cope with traffic fluctuations because of their anticipation capacity. Authors in \cite{zhang2017proactive} study NFV orchestration on line,and then estimate upcoming traffic and adjust VNF deployment a priori. There is also a hybrid scaling mechanism \cite{duan2017dynamic} that combines reactive and proactive scaling to improve system performance. Proactive scaling allocate resource in advance by predicting traffic, and adapt traffic splitting in actual operation reactively. 
 
 But there are still other problems. Current solutions mostly focus on how to predict traffic, rather than observing traffic characteristics in a specific NFV scenario. So, most of them use a uniform threshold to scale in/out. In real NFV scenario, each VNF may serve the one or more flows, and the characteristics of these flows are completely different, a uniform threshold used in this scenario is not suitable, because each VNF has a distinct processing logic depending on incident network traffic and events. Even if certain VNFs share packet processing functionality such as packet header analysis, the differences in upper-layer processing and implementation can exhibit unique resource usage patterns \cite{cao2017envi}.

We proposes a dynamic threshold scaling mechanism that can tailor thresholds according to each VNF's characteristic. As setting thresholds is a per-VNF task, and requires a deep understanding of workload trends and the diversity of each VNF, so we have added tailor-made features to the traditional dynamic mechanism. Besides, we also reserve resources by predicting workload and add an emergency module to cope with anomaly traffic, that is to say we develop a hybrid scaling policy combining proactive and reactive scaling together. Moreover, the sharp rise of network traffic not only can be caused by large amount of new incoming flows, but also can be induced by the growing of existing flows. If the traffic arises mainly due to the growing of existing flows, then only rerouting new flows can not alleviate the overload quickly and SLAs may be violated \cite{zhang2016co}. The only method to avoid SLA violations is to migrate flows and associated NF internal states quickly and safely from existing instances to new scaled instances, so state migration is an important part of the scaling procedure. We achieved the flow migration in scaling process on openNF \cite{zhang2016co} to guarantee the accuracy and timeline of scaling. 

To sum up, we have two main contributions:

First, we have developed a per-VNF scaling scheme based on dynamic threshold, who sets threshold for each VNF depending on their characteristics of services, traffic and resource utilization respectively. 

Second, we combine reactive and proactive scaling together to improve system performance, reactive scaling plays a role of emergency physician to handle some burst traffic while proactive scaling allocates resources in advanced according to traffic trends. 

We leverage RL algorithms, Deep Deterministic Policy Gradient (DDPG), to automatically learn the scaling policy, in more detail, we demonstrate the use of VNF resource utilization and traffic information as input features to train actor and critic network. OpenNF offers us an NFV control plane to guarantee the accuracy of flow migration during scaling in/out. We report preliminary experimental results on the caching proxy Squid and an asset detection and monitoring system,PRADS. 

In the remainder, the paper is organized as follows. We will discuss related work in Section II, and introduce our scheme in details in Section III. We then proceed to discuss the results of our experiments with others in Section IV. Finally, we conclude with Section V.  
\section{Motivation}

\section{Related Work}
The problem of scaling the NFV resource has been extensively studied and many auto-scaling techniques have been proposed by researchers, Considering the anticipation capacity as the main criteria, auto-scaling techniques could be divided into two main classes: \textbf{reactive} and \textbf{proactive}. Reactive scaling techniques are to add/remove VNF instances by responding to changes of runtime status of existing VNFs, and proactive policy is to predict traffic volumes based on the history \cite{duan2017dynamic}. Static threshold-based policies clearly belong to the reactive category, In contrast, machine learning, some dynamic threshold-based policies and some heuristic algorithm can be used with both reactive and proactive approaches. Whereas most current techniques are a hybrid scaling strategy, that combines reactive scaling and proactive scaling, such as \cite{delimitrou2014quasar}, they add traffic prediction function to the original reactive policies. In the present review, we will consider auto-scaling techniques grouped into three categories: \textbf{reactive scaling}, \textbf{proactive scaling} and \textbf{hybrid scaling}. 

\textbf{Reactive scaling}: Reactive scaling include threshold based rules and machine learning.Threshold based rules or policies can be divided into \textbf{static threshold} and \textbf{dynamic threshold}.The earliest solution is the static threshold-based solution,as the name suggest, predefined upper threshold and lower threshold are used for each performance metric. If the performance metric is above the upper threshold for a given period, scaling out action will be triggered \cite{tang2014efficient}. 
Cloud provider like Windows Azure \cite{bworld2} and AWS \cite{bworld1}\cite{bworld}, usually offer a threshold based auto-scaling mechanism, RightScale \cite{bworld3} propose combining regular reactive rules with a voting process. If a majority of the VMs agree on that they should scale out or scale in, that action is taken; otherwise, no action is planned. 

However, setting the suitable thresholds is a very tricky task, and may lead to instability in the system. Besides, static thresholds become invalid if the application behavior changes or traffic patterns are quite variable, which may need manual configuration and may be very  tricky and complex, thus, dynamic threshold-based scaling is born to solve the above problems. 

Little research has been done in the use of dynamic threshold, including the proportional threshold \cite{lim2009automated} or the adaptive thresholds introduced by Beloglazov and Buyya \cite{beloglazov2010adaptive}. Lim et al introduced a proportional threshold technology based on integral controller \cite{lim2009automated}. In \cite{beloglazov2010adaptive} authors collect the CPU utilization of each VM allocated in a host, then, they find out an interval of the CPU utilization, which will be reached with a low probability (for example, 90\% and 20\%), and therefore, set the corresponding thresholds. 

With the development of artificial intelligence, more and more researchers\cite{tang2014efficient}\cite{cao2017envi} have also applied\textbf{ machine learning }to the network. \cite{cao2017envi} used a neural network to achieve NFV auto-scaling, ENVI, periodically collects VNF-specific and infrastructure resource utilization information, and detects VNF scaling using this information and pre-trained machine learning models. 

Determined by the characteristics of machine learning, the accuracy of the model directly affects the accuracy of the scaling decision, thus, an inaccurate models can lead to large-scale system failure and may cause SLA violations. Even worse, However, the initial models may not perform well if VNFs undergo major software updates that affect their capacity and performance, in such cases, they have to retraining the initial neural network which waste a lot of time. On the contrary, our model has a high degree of universality and can be applied to many scenarios, no matter how the software updates, which is due to our agent is trained to deal with different situations. 


The main drawback of reactive techniques is that they not anticipate to unexpected changes in the workload, and therefore, resources cannot be provisioned in advance, a proactive scaling turned out to get over this drawback.

\textbf{Proactive approach:}\cite{zhang2017proactive} seek a proactive approach to provision new instances for overloaded VNFs ahead of time based on the estimated flow rates. They predict traffic and derive the processing capacity requested by VNFs. Then, simultaneously introduce vertical scaling by allocating adaptive processing capacities to the new instances when performing horizontal scaling (installation of VNF instances) ahead of time. Although, it considering the fluctuations of traffic traveling service chains and call for two other algorithms for new instances assignment and service chain rerouting, the accuracy of the two algorithms has yet to be improved and there is no scale-in mechanism to save cost. What's more, this model are too idealistic to take many actual factors into account,such as, constraints on the number of servers available or ramp constraints on the rerouting decisions. 


Proactive scaling can realize resource allocation in advance by predicting traffic, while reactive policy remedy prediction fault, therefore, a hybrid strategy exploits all opportunities for timely scaling of VNFs and significantly improves system performance. 

\textbf{Hybrid scaling mechanism:} The references \cite{duan2017dynamic}, ScalIMS caters to key control-plane and data-plane service chains in an IMS, combining proactive and reactive approaches for timely, cost-effective scaling of the service chains. it’s a hybrid scaling mechanism. When peak workloads arrive asynchronously across geographical spans, ScalIMS effectively reduces the total number of configured VNF instances by creating service chains across DCs, while guaranteeing outstanding QoS. But, the setting of the threshold in the reactive scaling is fixed, and which is inapplicable for real multi-traffic scenarios. 

Most of them never consider the complicated fluctuations of traffic volume, various traffic patterns and difference among servers. 
We target a approach combining proactive and reactive scaling, which is more practical given the time overhead for VNF deployment and considering the real link resource and server resource.

\section{DESIGN}
There are two main tasks included in the auto-scaling decision system: (i) decides the time to scale in or scale out, and (ii) chooses a proper number of instances to be launched or destroyed. The former task is realized reactively, because the scaling time is decided by observing whether the resource utilization exceeds the threshold value, while the latter is proactive, we reserve resources by predicting traffic volumes. We propose two algorithms to solve these two main problems in NFV scaling. The accurate scaling time is realized by DRL, a heuristic algorithm is designed to calculate the scaling amount of VNF instances.

\subsection{When to Scale} 
We have mentioned the importance of an accurate scaling time in the previous article:scale too early will lead to a waste of resources, and affect service quality while too late. When to scale is related to the resource utilization and user request characteristics of each VNF.Therefore, we set the threshold periodically according to the characteristics of each VNF: to scale out when its resource utilization beyond the high threshold, and to scale in while below the low threshold.DDPG is used to set up this equation and we call it DRL-threshold problem.We now formalize the DRL-threshold problem, and represent DRL-threshold as a discrete-time, continuous state and action space Markov decision processes (MDP).

\subsubsection{State/Input}

The input of this module is the VNF characteristic information, consist of type of VNF, flow protocol, traffic volume and resource utilization.  The characteristic information of each VNF can be represented by a list:

\begin{equation}
\begin{aligned}
VNFstate=\{F,M,S,L,Q,U,A\}
\end{aligned}
\label{1}
\end{equation}
\begin{itemize}
\item \textbf{The type of VNF(F):} There are significant differences between different types of VNF, and they may vary in the flow served, the packages processed, and the resource configuration properties.Besides, every VNF has a distinct processing logic depending on incident network traffic and events. Even if certain VNFs share packet processing functionality such as packet header analysis, the differences in upper-layer processing and implementation can exhibit unique resource usage patterns \cite{cao2017envi}.
\end{itemize}
\begin{itemize}
\item 
\textbf{The number of flow processed by the VNF(M):}  Moreover, the sharp rise of network traffic not only can be caused by large amount of new incoming flows, but also can be induced by the growing of existing flows. If the traffic arises mainly due to the growing of existing flows, then only rerouting new flows can not alleviate the overload quickly and SLAs may be violated[18]. The only method to avoid SLA violations is to migrate flows and associated NF internal states quickly and safely from existing instances to new scaled instances,Therefore, we need to know whether the increase in traffic is due to the new flows or existing flows, and then do the corresponding migration strategy (migrating new flows or existing flows). We judge this by observing the change in the number of VNF flows.
\end{itemize}

\begin{itemize}
\item \textbf{The number of packets processed by the VNF (S):} The number of packets processed by each VNF can directly reflect their current resource utilization, and periodic changes in traffic can be found by recording data packets at different times
\end{itemize}

\begin{itemize}
\item \textbf{the number of packets dropped by the VNF (L):}The packet loss rate can also reflect the current processing capacity of the VNF. A high packet loss rate means the current load of the VNF is heavy and may need to scale out. Therefore, the packet loss rate is of great reference value for setting the threshold.
\end{itemize}

\begin{itemize}
\item \textbf{the queue length of the VNF (Q):} Queuing takes into account the processing capacity of VNF. Even if there is no packet loss or the loss rate is low, if the queue is very long, it means the load of VNF is heavy and may be about to drop packets.
\end{itemize}

\begin{itemize}
\item \textbf{the cpu utilization of the VNF (U):} CPU utilization is the most important parameter that directly reflects the relationship between current traffic and VNF processing capacity.So the threshold we set, namely, the output of deep learning, is also related to the CPU utilization of VNF.
\end{itemize}

\begin{itemize}
\item \textbf{Scale policy: scale in/out or do nothing (A):} In order to prevent frequent scale, we should refer to scale action in the last cycle.so that we can remind the agent, If some other inputs: packet loss or Ql are high, and there is no scaling out in last cycle, we may need to adjust its threshold value, in order to scale out to reduce packet loss as soon as possible, but if the VNF had just scaled out,we should observe for a while.
\end{itemize}

\subsubsection{Action/Output and Reward}

The output is the threshold vector of CPU utilization, calculated by DDPG based on current status information:
\textit{upper threshold,lower threshold}.
The two objectives of our model are maximizing service performance while minimizing the costs. Thus, we craft a reward function according to the two objectives.

During each training step, a reward signal Rwd is set to reflect the performance of the scaling decisions, which is expected not only to ensure that the minimum number of VMs are used to provide the fastest processing, but also to ensure the accuracy and stability of the system. Hence, we might extend the function to take packet loss and queue length into consideration. The above intuition can be expressed as follows:

\begin{equation}
\begin{aligned}
R = \frac{\sum_{i=1}^N(\sum_{j=1}^Mv_{i,j}-{L_i}-{Q_{i}})\ }{N}
\end{aligned}
\label{2}
\end{equation}

where $v_{i,j}$ is the traffic volume of flow j processed by VNF i, the first term in the equation (1) shows the average throughput on each VNF, obviously, the more traffic process, the less VM consume, and the more rewards we get. $L_i$ is the number of packet loss in last unit time on VNF i, and the second term in the equation is intend to ensure accuracy of our model. $ Q_i$ is the average packet queue length of VNF i, and the third term means the longer the queue, the less reward, which help us further consider future traffic demand to improve stability and sensitivity.

In short word, the first term in function encourage our model to pursue a more efficient and low cost goal, and latter two ensure SLA to meet a balance by punishing some poor situation. 

\subsection{How Many VNFs need to reserve}
Now we have got the appropriate threshold,calculated by DDPG, and we can determine which VNF need to scale in/out.
But, it takes a long time to turn on the VM (as mentioned above), thus, we may need to keep a certain number of VMS on before we decide to scale out, so as to ensure a rapid implementation of flow migration.
but how many idle VMs is required is another tricky question, insufficient amount of VMs can not alleviate the phenomenon of overload, but too much will lead to waste resources, so we design a heuristic algorithm to calculate the reasonable amounts to scale.
 
First of all, we need to introduce an important concept:two buffer queues which contain all VM resources, one consists of all running VM queues, and the other is idle VM queue. 
Their relationship can be expressed by the following formula:

\begin{equation}
\begin{aligned}
N = N_{run} +N_{idle}    
\end{aligned}
\label{3}
\end{equation}

Which means the total VMs N are consist of $N_{run}$ running VMs and $N_{idle}$ idle VMs.

The idle VM queue is an ordered queue. We refer to the running queue as the resource utilization queue,
the other idle VM queues $N_{idle}$ are called the standby queues, ranked in chronological order. As VMs reboot takes a long time,whenever we scale in, there are some VMs need to be removed, instead of directly shutting it down, we enqueue it to the tail of the respective idle buffer VM queues $N_{idle}$ and keep them on for some cycles to avoid frequent creation or release of VNF instances under traffic fluctuations. How many circles is customizable. Once a VNF instance is enqueued, no more flows will be routed to it. Whenever there is a VNF need to scale out, if there are available VMs in the buffer queue of this VNF, buffered VMs will be popped out from the tail of the queue, and transformed back to working VMs, to fulfill the demand as much as possible. Then, we can migrate flows on these open VM directly, don’t need to restart the new VM, such not only can save the cost and time, but also avoid oscillations caused by frequent startup of the virtual machine.
  
There are also studies on buffer queues in other related work \cite{fei2018adaptive}\cite{duan2017dynamic}, but they point that a VNF instance will stay put once it is created in a scaling interval $t$, then the algorithm will tag it with $t$ and enqueue it to the tail of the queue. In under-provisioning case, buffered instances from the tail of the respective queue will be popped out and serve for those requested service chains. Unused VNF instances in the queue are removed after $\kappa$ time slots, which can be tuned by the NFV provider. 

But we think there are some problems with this, first of all, every VNF is located differently, whenever a VNF needs to be scaled out, we must first consider the distance between it and its migrating object when selecting the object VNF to be migrated, that is, we choose an existing VNF instance with minimum transportation cost, which is not considered in \cite{fei2018adaptive}\cite{duan2017dynamic}.In \cite{fei2018adaptive}\cite{duan2017dynamic}, what they more concentrate on is how to avoid oscillations, thus, the longest remaining VNF in the queue will be picked. 

Secondly, what they reserve is VNFs, but we reserve VMs. That is to say, for their schemes, the VNF reserved can only be used when the same VNF need to scale out. Otherwise, even if other types of VNF need to be expanded, the reserved VNF cannot back up. Inversely, we reserve VM not VNF, so as to serve all kinds of VNF expansion requests, and improve the flexibility of resource significantly. what's more, we are considering different time cost to choose reserve resources. They keep VNF as backup to save VNF replication time, while we aim at saving start time of VMs, however, it's much faster to copy a VNF than to start a VM, therefore, keep VM as backup resources is more effective.

Thirdly, the way in which they reserve idle VNFs as reserved resources also lacks certain considerations, because these reserved idle VNFs consume costs as other running VNFs, and affects the utilization of resources in our system. Therefore, in this paper, the management of these idle VNFs is studied in detail. The retention time of each VNF is related to the current overall resource utilization and energy efficiency ratio.

We first introduce a concept, energy efficiency ratio, $\gamma$,which represents the ratio of processed traffic $V$ to current computation resource $C$, the compute capability of VM is proportional to their CPU utilization,$C=f(U)$,we assume that $C=f(U)=H*T$,H means the upper threshold of CPU utilization and T is  the unit time, which can entirely fit resource consuming rate. The product of the above two represents the computing capacity of each virtual machine per unit time in the current system. 

$\gamma^*$  represents the ideal energy efficiency ratio, which can be tuned by the NFV provider. $\gamma_{run}$ represents the energy efficiency ratio in actual operation and can be calculate by the following equation (4):

\begin{equation}
\begin{aligned}
\gamma_{run}& = \frac{V}{C}\\
&=\frac{\sum_{i=1}^{N_{run}}\sum_{j=1}^{M}\int_0^T v_{i,j}\,dt}{\sum_{i=1}^{N}{H*T}}\\
&=\frac{\sum_{i=1}^{N_{run}}\sum_{j=1}^{M}\int_0^T v_{i,j}\,dt}{NHT}\\
\end{aligned}
\end{equation}

  In which, $N_{run} $ represents the number of running VNF while  $N_{idle}$ represents the number of idle VNF in the buffer queue. 
  The numerator is total traffic volume,and the denominator represents two part resource of the running VNF and idle VNF, respectively. 
  
  The preferable scenario is $\gamma_{run}==\gamma^*$, so we hope to adjust $N_{idle}$ to bring the energy efficiency ratio $\gamma_{run}$ closer to the ideal energy efficiency ratio. So, a best fit number of idle VNF under ideal circumstances, $N_{tuned}$, will be calculated in the following way:

\begin{equation}
\begin{aligned}
N_{tuned}=\frac{\gamma_(N_{run}+N_{idle})-\gamma^*N_{run}}{\gamma^*}  
\end{aligned}
\end{equation}

$N_{tuned}$ is the theoretically optimal amount of idle VNFs, but, we have to replace it with zero when it's minus,which means the loads of all running VNFs are light and there is no need to have idle VNF as a backup, so we will turn off the VNF in the idle buffer queues to save resources.
\begin{equation}
 N_{idle}^*=\left\{
\begin{aligned}
&N_{tuned} &   & N_{tuned}\geqslant 0 \\ 
&0 &  &else \\
\end{aligned}
\right.
\end{equation}

\begin{table}
\linespread{1.5}

\caption{Mathematical notations}
\label{tab:1}       
%
\begin{tabular}{lll}     
\hline\noalign{\smallskip}       
\multicolumn{2}{l}{\textbf{\emph{State parameters of VNF}}}  \\  
\noalign{\smallskip}\hline\noalign{\smallskip}
F&The type of VNF \\
\noalign{\smallskip}
M& The number of flow processed by the VNF \\
\noalign{\smallskip}
S&The number of packets processed by the VNF \\ 
\noalign{\smallskip}
L&The number of packets dropped by the VNF  \\
\noalign{\smallskip}
Q&The queue length of the VNF \\ 
\noalign{\smallskip}
U&The CPU utilization of the VNF \\ 
\noalign{\smallskip}
A& Scale policy: scale in/out or do nothing \\ 
\noalign{\smallskip}
H& The upper threshold \\ 
\noalign{\smallskip}
D& The lower threshold \\ 
\noalign{\smallskip}
\hline\noalign{\smallskip}  
\multicolumn{2}{l}{\textbf{\emph{Formula variable}}}  \\  
\noalign{\smallskip}\hline\noalign{\smallskip}
R& The reward of machine learning  \\
\noalign{\smallskip}
$v_{i,j}$ & The traffic volume of flow j processed by \\&VNF i, i $\in$ N, j $\in$ M   \\
\noalign{\smallskip}
$L_i$ & The number of packets dropped by the VNF i   \\
\noalign{\smallskip}
$Q_i$ &The queue length of the VNF i \\ 
\noalign{\smallskip}
N& The total number of VNF  \\
\noalign{\smallskip}
$N_{run}$ & The number of running VNF  \\
\noalign{\smallskip}
$N_{idle}$ & The number of idle VNF  \\
\noalign{\smallskip}
$\gamma$ & Energy efficiency ratio:  the ratio of processed \\& traffic to consumed CPU resources \\
\noalign{\smallskip}
$\gamma_{run}$ & The energy efficiency ratio in
actual operation \\
\noalign{\smallskip}
$\gamma^*$ & The ideal energy efficiency ratio, up to provider \\
\noalign{\smallskip}
V & The sum of traffic volume handled by VNF\\
\noalign{\smallskip}
C & Resource consumed\\
\noalign{\smallskip}
$c_i$ & Resource consumed by VNF i\\
\noalign{\smallskip}
\noalign{\smallskip}\hline
\end{tabular}
\end{table}

\section{Implementation}
We implement our model on openNF \cite{gember-jacobson2015opennf}, a control plane architecture that provides efficient, coordinated control of both internal NF state and network forwarding state to allow quick, safe, and fine-grained reallocation of flows across NF instances.Using OpenNF,we can achieve  accurate flow migration during scaling process.For detailed openNF information, please refer to Reference \cite{gember-jacobson2015opennf}.We can collect the status information of VNF and publish the threshold value through openNF.The overview architecture is shown in figure 2.
\begin{figure}[tbp]
\centering
\includegraphics[width=70mm,height =65mm]{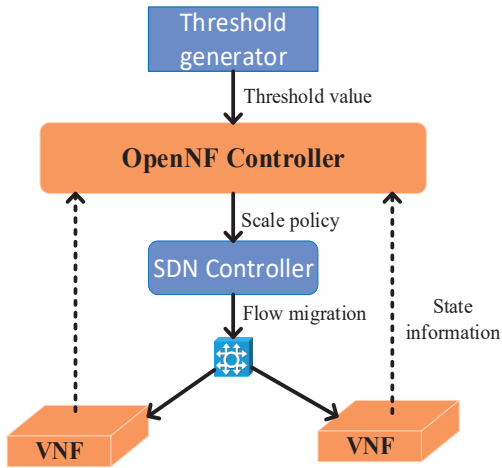}   
\caption{architecture.}
\label{fig.2}
\end{figure}

To get a good understanding of our system,we divide our model into four parts as Figure 3: a resource monitor (RM), a training model for deep reinforcement learning (DDPG), an emergency processing module (EP), and an auto-scaling engine (ASE).We implemented three modules: VNF monitor, emergency engine and scaling engine in openNF controller. Threshold generator was realized outside openNF. Threshold generator and openNF were implemented on different machines, we simply use UNIX socket to exchange information.Please refer to figure 2 and 3 to have a better understanding of our system.

\begin{figure}[tbp]
\centering
\includegraphics[width=65mm,height =60mm]{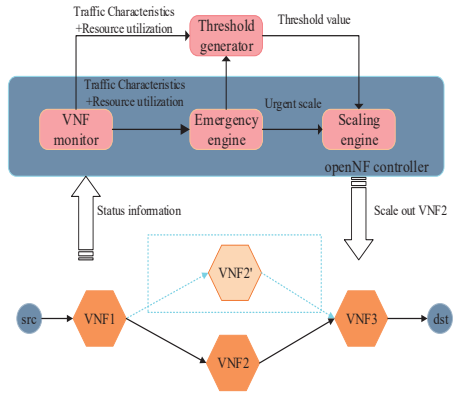}   
\caption{workflow}
\label{fig.1}
\end{figure}

The specific functions of each module are as follows:

\subsection{Resource Monitor}

Resource Monitor (RM) can collect the information of all VNFs, including, the number of flows, link utilization, and other relevant network metrics to upload to the other two modules: DDPG and ASE. It continuous monitors, but periodic uploads, except for special cases, such as a sudden serious SLA violation, which need to be handled at once, under the circumstances, RM will immediately send the information to EP module for alarm.

 \subsection{Threshold Generator}
In threshold generator, we leverage RL algorithms to automatically learn the status information of VNF and output scaling thresholds. There are three types of classical RL algorithms that are commonly used to train the control policies: Q-learning, policy gradient, and actor-critic algorithms. These algorithms typically assume that the action space is discrete, which simplifies the search for actions. However, in our design, it is more natural to consider a continuous threshold value as the decision. In this case, recently proposed deterministic policy gradient algorithm is more desirable and allow one to directly learn and enact a deterministic instead of a stochastic policy. Thus, we leverage the Deep Deterministic Policy Gradient (DDPG) \cite{lillicrap2016continuous} algorithm to train actor and critic network and we implement it via Tensorflow, which is a professional and efficient data mining and data analysis tool.

\subsection{Emergency Processing}
Although our model will have rich experience in training to deal with the various states of the current VNF, some unpredictable conditions will still appear in the system, in order to prevent system performance degradation due to some sudden factors, or in order to improve the robustness of the system, we add an emergency processing(EP): module to handle some unexpected but serve incidents, for example: a sudden increase in packet loss rate. Assuming that there is no EP, and the packet loss happen to occur at the beginning of a new cycle, the monitor have to wait, until the end of this cycle, to upload alarm to DDPG, by then, DDPG will react to this emergency which may have already caused a serious violation of SLA. The EP module is designed in consideration of the above situation.

In a normal situation, even if there is a surge in traffic, but it does not affect the SLA, then the monitor will still just report the information to DDPG for regular processing, which is periodic. But if the monitor detects that the system has severe SLA violations, like a sudden increase in the queue length, or a sudden increase in the packet loss rate, etc. the monitor will immediately inform EP module for emergency scaling, which is timely, not periodic. EP module pays direct attention to the performance of the service, avoid the mistakes caused by inaccurate training in machine learning and some unpredictable and unreasonable system bugs.

\subsection{Auto-Scaling Engine}
ASE contains two buffer queues, we have mentioned above, to calculate the appropriate the amount of resources allocated. It's important to emphasize that they are just two logical buffer queues to temporarily tag every VNF instances for each type of VNF. The logical queue is not a real queue that occupies any specific server node \cite{duan2017dynamic}.
The main work of scaling engine is carry on scaling based on a comparison between current statue and threshold. When scaling out, we select an “closest” VNF in the idle VNF queue while, scaling in, a lightest loaded VNF will be picked to undertake the rest of the traffic of the VNF which needs to be removed soon. The workflow flowchart is shown in figure 3.

\textbf{Traffic patterns.} As we want to simulate real traffic scenarios as much as possible, we observed several typical traffic scenarios with user characteristics as our experimental scenario:

\begin{itemize}
\item The traffic is flat and generally periodic, which is in line with users' usage habits, with peaks and valleys, As figure 3 shows.
\end{itemize}
\begin{itemize}
\item The traffic is highly variable and generally periodic, and sometimes there will be two or three spikes in traffic, which may be accompanied by some hot events,just like fig 4 shows:
\end{itemize}
\begin{itemize}
\item The flow changes frequently, and it fluctuates greatly, without periodicity. This situation is more in line with................,which is shown in figure 5
\end{itemize}

Assuming that the flow size is almost the same, we change the overall traffic volume by controlling the arrival time of every flow and traffic rate to simulate the above three scenarios.

\section{Evaluation}
In this section, we compare our scheme with static threshold policy under the above three traffic patterns, we select two VNFs as objects of observations: a caching proxy Squid and iptables, a userspace command line program that facilitates configuration of packet filtering rules in the Linux kernel to realize firewall and network address translator functionality. And We evaluate the performance of them based on three groups of metrics:
\begin{itemize}
\item Packet loss rate, which can reflect the reliability of the scaling scheme.Due to static threshold policy passivity, during the boot-up time of new instances, traffic continues to arrive at the overloaded VNF instances, resulting in a high packet loss rate and then high RTT.
\end{itemize}

\begin{itemize}
\item Considering the trade-off between resource utilization and SLA performance of auto-scaling scheme, we customize $\alpha = flow_{t}*N $ as a standard to evaluate two schemes, where t means the average flow completion time, and N represents the total number of host consumed to process all the traffic. We generated the same traffic for the two schemes, and we expected the $\alpha$ to be as low as possible, which means that we spent only a smaller amount of VM resources, but the packet was processed very quickly, that is, the given traffic was processed in a shorter time.
\end{itemize}

\begin{itemize}
\item $Pun = ltc * N$ is another punishment parameters, ltc represents the average latency of all packets processed by this VNF, which will directly lead to SLA violations if its value is too large. Similarly, too many virtual machines also mean a waste of resources, thus we use the product of the above two parameters to achieve our goal:shorter packet latency with fewer virtual machine resources.
\end{itemize}

\subsection{Squid}
Figure 1 shows the results of a comparison experiment on Squid observed in the first traffic pattern.

Figure 2 shows the results of a comparison experiment on Squid observed in the second traffic pattern.
\subsection{iptables}
Figure 4 shows the results of a comparison experiment on iptables observed in the first traffic pattern.
\cite{gember-jacobson2015opennf}
Figure 5 shows the results of a comparison experiment on iptables observed in the second traffic pattern.
\bibliographystyle{plain}
\bibliography{reference}

\end{document}